\documentstyle[preprint,aps]{revtex}
\begin{document}
\draft

\widetext
\title{ Effective lattice actions for correlated electrons  }
\author { Antimo Angelucci }
\address {Institut f\"ur Theoretische Physik, Universit\"at W\"urzburg,
D--97074 W\"urzburg, Germany \\
          Institut f\"ur Physik, Universit\"at Augsburg,
D--86135 Augsburg, Germany$^*$}
\date{\today}
\maketitle
\begin{abstract}

We present an exact, unconstrained representation of the electron
operators in terms of operators of opposite statistics. We propose a
path--integral representation for the $t$-$J$ model and introduce a
parameter controlling the semiclassical behaviour. We extend the
functional approach to the Hubbard model and show that the mean--field
theory is equivalent to considering, at Hamiltonian level, the
Falikov--Kimball model. Connections with a bond-charge model are also
discussed.

\end{abstract}

\narrowtext

The investigation of the nature of the ground--state and of the low--lying
excitations of strongly correlated electron systems is a fundamental
problem of the modern many--body theory. The standard model of correlated
electrons on a lattice is the Hubbard Hamiltonian
\begin{equation}
H_H =  -t\sum \limits_{i,j,s } \Lambda_{ij}
c^{\dag}_{i,s} c_{j,s}   +
U \sum \limits_{i} n_{i,\uparrow}
                   n_{i,\downarrow},
\label{hubbard}
\end{equation}
where $\Lambda_{ij}$ is the symmetric adherence matrix connecting the
nearest neighbor (n.n.) sites of a hypercubic lattice of $M$ sites,
$c^{\dag}_{i,s}$ and $c_{i,s}$ are creation and annihilation operators
for electrons with spin projection $s=\uparrow,\downarrow$, respectively,
and $n_{i,s}= c^{\dag}_{i,s}c_{i,s}$.
Henceforth we denote the states at site $i$ with the notation
\begin{eqnarray}
  \vert 1, \!\Uparrow\,   \rangle_i & = &
  c^{\dag}_{i,\uparrow}   \vert \nu \rangle,  \quad\quad
  \vert 0, \!\Uparrow\,   \rangle_i  =  \vert \nu \rangle_i, \nonumber \\
  \vert 1, \!\Downarrow\, \rangle_i & = &
  c^{\dag}_{i,\downarrow}  \vert \nu \rangle, \quad\quad
  \vert 0, \!\Downarrow\, \rangle_i =
  c^{\dag}_{i,\downarrow}c^{\dag}_{i,\uparrow} \vert \nu \rangle,
\label{states_el}
\end{eqnarray}
where $\vert \nu \rangle = \bigotimes_i \vert \nu \rangle_i$.
The essential behaviour of $H_H$ in the strong-coupling regime
${U/{\vert t \vert}} \gg 1$ below half--filling is commonly
investigated\cite{vollhardt} by means of the effective Hamiltonian
$H^{\rm eff} = H_{tJ} + H^{(3)}$ acting in the subspace without doubly
occupied states $\vert 0, \!\Downarrow\, \rangle_i$, where
\begin{equation}
H_{tJ} =  -t \sum \limits_{i,j,s } \Lambda_{ij}
{\tilde c}^{\dag}_{i,s} {\tilde c}_{j,s}  +
          J \sum \limits_{\langle i,j\rangle }
(\vec S_i \, \vec S_j - {1\over 4 } N_i \, N_j),
\label{tj}
\end{equation}
is the $t$-$J$ model and $H^{(3)}$, not explicitly displayed\cite{vollhardt},
is the three--site term. Here $\langle i,j \rangle$ denotes the
summation over the n.n. sites, $J=4t^2/U$,
${\tilde c}_{i,s} =  c_{i,s}(1-n_{i,{\bar s}})$, with
${\bar s}$ denoting the opposite spin projection, are projected electron
operators, and
$\vec S_i = {1\over2}\sum_{s,p}
c^{\dag}_{i,s}{\vec \sigma}_{sp}c_{i,p}$ and
$N_i=\sum_{s}{\tilde c}^{\dag}_{i,s}{\tilde c}_{i,s}$ are the
spin and number operators, respectively, with
${\vec \sigma}_{sp}$ the Pauli matrices.

To approach the problem of strong correlation, slave--boson (--fermion)
methods\cite{vollhardt,zou} have been widely employed, because
they give a physically intuitive way to work in the subspace without
doubly occupied states and allow the introduction of mean--field approximations
by assuming condensation of the bosons. In the present investigation we
provide a path--integral description of the models (\ref{hubbard}) and
(\ref{tj}), starting from
the observation that the decomposition of the electron operators
via operators of opposite statistics can be achieved in terms of
exact operator identities.
Hence, contrary to the slave--particle methods, in our approach constraints
among the operators are absent.
In fact, the states (\ref{states_el}) can be
generated by means of canonical spinless fermions $f_i^{\dag}$
and Pauli operators $\vec\sigma_i$
\begin{eqnarray}
  \vert 1, \!\Uparrow\,   \rangle_i & = &
  \vert v   \rangle_i,   \quad\quad\quad\,\,\,
  \vert 0, \!\Uparrow\, \rangle_i  =
  f_i^{\dag}  \vert v   \rangle,   \nonumber \\
  \vert 1, \!\Downarrow\, \rangle_i & = &
  \sigma_{i,-}  \vert v   \rangle, \quad\quad
  \vert 0, \!\Downarrow\, \rangle_i =
  f_i^{\dag} \sigma_{i,-} \vert v   \rangle,  \label{states}
\end{eqnarray}
where $\sigma_{i,\pm} = (\sigma_{i,x}\pm i\sigma_{i,y})/2$ and
$\vert v \rangle = \bigotimes_i\vert v \rangle_i$ is the reference vacuum
annihilated by the $f_i$'s and $\sigma_{i,+}$'s.
We observe that the representation is closed, since no new states can be
generated by applying  $f_i^{\dag }$ and/or $\sigma_{i,-}$, so that
the operators of a given basis can be expressed in terms of the others
\begin{eqnarray}
&& f_i^{\dag}   = {\bar n}_{i,\downarrow} \,c_{i,\uparrow} -
                  n_{i,\downarrow}\, c^{\dag}_{i,\uparrow}, \quad
   \sigma_{i,-}  = c^{\dag}_{i,\downarrow} \, (c_{i,\uparrow} +
               c^{\dag}_{i,\uparrow}),  \label{map} \\
&& c^{\dagger}_{i,\uparrow}  =  \gamma_{i,+}f_i - \gamma_{i,-}f_i^{\dag}, \quad
\,\,\,\,\,\,
               c^{\dagger}_{i,\downarrow}   =  \sigma_{i,-}\,(f_i+f^{\dag}_i),
\label{op_electr}
\end{eqnarray}
where ${\bar n}_{i,s}=1-n_{i,s}$ and
$\gamma_{i,\pm} = (1\pm \sigma_{i,z})/2$. Hence, it is possible to
construct nonlinear combinations of the electron operators satisfying to
$\{f_i^{\dag},f_j\} = \delta_{ij}$,
$[\sigma_{i,\alpha},\sigma_{j,\beta}] = 2i\,\delta_{ij}
         \epsilon^{\alpha\beta\gamma} \sigma_{j,\gamma}$
or, relevant for us, one sees that the anticommutation relations
$\{c_{i,s}^{\dag},c_{j,p}\} = \delta_{ij}\delta_{sp}$
can be fulfilled via operators of opposite statistics.

For the moment, it is useful to avoid referance to a particular Fock
representation. To this aim we say that the
model (\ref{hubbard}) [or (\ref{tj})] acts on ``objects''\cite{sutherland}
$\vert \xi,\sigma \rangle_i$ of four (or three) different species,
placed exactly one to a site, which we label by defining the ``grade''
$\pi(\xi)=(-)^{\xi}=\pm1$ ($\xi=0,1$), i.e., by dividing the species
into even and odd, and the ``isospin'' $q_z=\pm{1\over2}$
(pictorially $\sigma=\Uparrow,\Downarrow$) quantum numbers.
The unitary transformations of the species assigned to
each site are given by Hubbard operators\cite{vollhardt,zou}
$X^{\xi\sigma}_{i{\xi\prime}{\sigma\prime}}   = \vert \xi , \sigma\rangle_i
   \langle {\xi\prime} , {\sigma\prime} \vert_i $,
which form the fundamental representation of the su($2\vert 2$) graded
algebra\cite{vollhardt,essler}. The special character is due to the
completeness relation $\sum_{\xi\sigma} {N}_{i,\xi\sigma} = 1$, where
${N}_{i,\xi\sigma}=X^{\xi\sigma}_{i\xi\sigma}$ are the local densities,
expressing the one--to--one correspondence between objects and sites.
With our notation, the anticommuting, or ``odd'' operators, are those
changing the grade [i.e., with $\pi(\xi)=-\pi(\xi\prime)$], henceforth denoted
\begin{mathletters}
\begin{equation}
\FL
\chi_{i,\sigma} =
X^{0\Uparrow}_{i1\sigma},  \,\,
   \chi_i^{\sigma} = X^{1\sigma}_{i0\Uparrow},  \,\,
\tau_{i,\sigma} =
X^{0\Downarrow}_{i1\sigma},  \,\,
\tau_i^{\sigma} = X^{1\sigma}_{i0\Downarrow}, \label{hub-odd}
\end{equation}
so that $\chi_i^{\sigma} = \chi_{i,\sigma}^{\dag}$, and
$\tau_i^{\sigma} = \tau_{i,\sigma}^{\dag}$.
For the commuting, or ``even'' operators,
[i.e., those with $\pi(\xi)=\pi(\xi\prime)$] we define the linear combinations
\begin{eqnarray}
S_{i,z} & = & {1\over2} X^{1\Uparrow}_{i1\Uparrow}  -
              {1\over2} X^{1\Downarrow}_{i1\Downarrow},
\quad\,\, S_{i,+} = X^{1\Uparrow}_{i1\Downarrow},  \nonumber \\
S_{i,t} & = & {1\over2} X^{1\Uparrow}_{i1\Uparrow}  +
              {1\over2} X^{1\Downarrow}_{i1\Downarrow},
\quad\,\, S_{i,-} = X^{1\Downarrow}_{i1\Uparrow}, \label{hub-spin}\\
L_{i,z} & = & {1\over2} X^{0\Uparrow}_{i0\Uparrow}  -
              {1\over2} X^{0\Downarrow}_{i0\Downarrow},
\quad\,\, L_{i,+} = X^{0\Uparrow}_{i0\Downarrow},  \nonumber \\
L_{i,t} & = & {1\over2} X^{0\Uparrow}_{i0\Uparrow}  +
          {1\over2} X^{0\Downarrow}_{i0\Downarrow},
\quad\,\, L_{i,-} = X^{0\Downarrow}_{i0\Uparrow}. \label{hub-pseudospin}
\end{eqnarray}
\end{mathletters}
For graded (i.e., supersymmetric) algebras, the grading of the states is a
convention of purely formal nature, because the grading of the operators
does not depend on the choice used. Physically, this freedom corresponds to
the fact that the {\it particle--hole} transformation
$\vert 0, \sigma \rangle_i \,\leftrightarrow\, \vert 1, \sigma \rangle_i$
leaves su(2$\vert$2) unaltered, i.e.,
\begin{eqnarray}
(S_{i,t},{\vec S}_i) & \leftrightarrow & (L_{i,t},{\vec L}_i), \nonumber
\\
(\chi^{\Uparrow}_i, \chi^{\Downarrow}_i, \tau^{\Uparrow}_i,
  \tau^{\Downarrow}_i) & \leftrightarrow &
  (\chi_{i,\Uparrow}, \tau_{i,\Uparrow},
  \chi_{i,\Downarrow}, \tau_{i,\Downarrow}),
\label{p-h}
\end{eqnarray}
where $S_{i,\pm}=S_{i,x}\pm iS_{i,y}$ and $L_{i,\pm}=L_{i,x}\pm iL_{i,y}$.
Hereafter we shall always use the
correspondence (\ref{states_el}), so that  ${\vec S}_i$ and ${\vec L}_i$ are
identified as the local spin and pseudospin operators, respectively.
Henceforth we also define the isospin vector as the sum of spin and
pseudospin: ${\vec Q}_i=\vec S_i + \vec L_i$.
The total numbers of species-$(\xi,\sigma)$ objects
$N_{\xi\sigma} = \sum_i N_{i,\xi\sigma}$ are related to
the total numbers of spin-$s$ electrons, henceforth denoted
$N_{s}$, through the self-evident relations
$N_{\uparrow} = N_{1\Uparrow} + N_{0\Downarrow}$ and
$N_{\downarrow} = N_{1\Downarrow} + N_{0\Downarrow}$,
from which we obtain the equivalent expressions
\begin{eqnarray}
S_{z} & = & {1\over2}(N_{1\Uparrow} - N_{1\Downarrow}), \quad
S_{z}   =   {1\over2}(N_{\uparrow} - N_{\downarrow}),  \nonumber \\
L_{z} & = & {1\over2}(N_{0\Uparrow} - N_{0\Downarrow}), \quad
L_{z}   =   {1\over2}(M - N_{\uparrow} - N_{\downarrow}), \label{conservation}
\end{eqnarray}
where ${\vec S}=\sum_i {\vec S}_i$ and ${\vec L}=\sum_i {\vec L}_i$.
The operators $S_z$ and $L_z$ are conserved because $N_\uparrow$ and
$N_\downarrow$ are constant, however from Eq.\ (\ref{conservation})
we might also say that conservation follows because even and odd
objects, when not conserved, are both created and/or destroyed in pairs.
The latter is a useful
way to visualize the less intuitive properties of the pseudospin.
The picture is sharpened by observing that spin and pseudospin always
act on $N_{\xi\sigma}$ as raising--lowering operators, thus changing
$N_{\xi\sigma}$ by one unit in states with a definite number of objects.
We have  $ \left[N_{0\Uparrow}, L_{\pm} \right] = \pm L_{\pm} $, and
$ \left[N_{0\Downarrow}, L_{\pm} \right] = \mp L_{\pm}$,
as well as the commutators where
$N_{0\sigma} \to N_{1\sigma}$ and
$L_{\pm} \to S_{\pm}$.

Hubbard operators are useful for discussing symmetry properties and by using
them it is easily seen that {\it any} transformation
of the electron operators can be exactly rephrased in the basis (\ref{states}).
In particular, in the representations (\ref{states_el}) and (\ref{states})
the particle--hole transformation (\ref{p-h}) is achieved by
letting $c^{\dag}_{i,\uparrow} \leftrightarrow \, c_{i,\uparrow}$ and
$f_i^{\dag} \leftrightarrow \, f_i$, respectively.
The realization of Eq.\ (7) in the basis (\ref{states}) is
\begin{eqnarray}
\FL
\chi_i^{\Uparrow} & = &  \gamma_{i,+}f_i,  \,\,\,\,
\tau_i^{\Uparrow} = \sigma_{i,+}f_i, \,\,\,\,
S_{i,t} = {1\over2} (1-n_i), \nonumber \\
\FL
\chi_i^{\Downarrow} & = & \sigma_{i,-} f_i,  \,\,\,\,
\tau_i^{\Downarrow} = \gamma_{i,-} f_i, \,\,\,\,
{\vec S}_i = {1\over2}{\vec \sigma}_i (1-n_i),
\label{hubbard_sf}
\end{eqnarray}
where $n_i=f_i^{\dag}f_i$, and the operators not explicitly displayed
are easily obtained by means of Eq.\ (\ref{p-h}). The local densities are
$N_{i,0\Uparrow}=\gamma_{i,+}n_i$,
$N_{i,0\Downarrow}=\gamma_{i,-}n_i$,
$N_{i,1\Uparrow}=\gamma_{i,+}(1-n_i)$,
$N_{i,1\Downarrow}=\gamma_{i,-}(1-n_i)$, and automatically satisfy
to $\sum_{\xi\sigma} {N}_{i,\xi\sigma} = 1$. For the realization of Eq.\ (7)
with electron operators we refer instead to Ref.\cite{essler,schad}.

We now consider  the auxiliary model\cite{last}
acting by permutation of the four species of objects
$\vert \xi,\sigma \rangle_i$
\begin{equation}
H^{(2\vert 2)} =   \sum \limits_{\langle i,j \rangle} \left[
        - t P_{ij}^{01} + {J\over2}(P_{ij}^{11} - P_{ij}^{00})\right],
\label{kor}
\end{equation}
where $P_{ij}^{11} = 2({\vec S}_i{\vec S}_j + S_{i,t}S_{j,t})$ and
$P_{ij}^{00} = 2({\vec L}_i{\vec L}_j + L_{i,t}L_{j,t})$ are the spin and
pseudospin permutators acting nontrivially on the odd and even objects,
respectively, and $P_{ij}^{01} =
\chi_i^{\sigma} \chi_{j,\sigma}  + \tau_i^{\sigma} \tau_{j,\sigma} + h.c.\,$
is the operator permuting pairs of objects of opposite grade.
Eq.\ (\ref{kor}), henceforth referred to as the extended model,
is the case $x=y=2$ of the class of Hamiltonian $H^{(y\vert x)}$
permuting objects of $y$ odd and $x$ even different species,
introduced by Sutherland\cite{sutherland} for the special case
$J=\pm 2\vert t \vert$ in studying exactly solvable systems in one dimension.
The spectrum of $H^{(2\vert 2)}$ contains that of $H^{(2\vert 1)}$ by
construction and noting that the latter is actually the $t$-$J$ model,
we see that information about the Hamiltonian (\ref{tj}) can be obtained
from the model (\ref{kor}) by imposing  the
{\it conserved} constraint $N_{0\Downarrow} = 0$.
Using Eq.\ (\ref{hubbard_sf}) $H^{(2\vert 2)}$ becomes
\begin{equation}
H_{tJ}^{\rm ex} =   t \sum \limits_{i,j} \Lambda_{ij} P_{ij} f_i^{\dag} f_j +
{J\over2} \sum \limits_{\langle i,j \rangle}
         \Delta_{ij}(P_{ij} - 1),
\label{tjeff}
\end{equation}
where $P_{ij} = {1\over 2}(1 + \vec \sigma_i \, \vec \sigma_j)$ and
$\Delta_{ij} = (1 - n_i - n_j)$, and where we have added a constant
to ensure that Eq.\ (\ref{tjeff}) reduces to the standard definition (\ref{tj})
of the $t$-$J$ model for $N_{0\Downarrow} = 0$.
We consider the generalization (\ref{kor}) useful\cite{three} because it leads
in $H_{tJ}^{\rm ex}$ both {\it i)} to the isotropic contribution
$P_{ij}$ in the hopping term,
similarly to the generalization of Khaliullin\cite{Kha},
and {\it ii)} to the quadratic contribution
$\Delta_{ij}$ in the magnetic term. In this respect,
Eq.\ (\ref{tjeff}) is easier to study than the original $t$-$J$ model,
as for, e.g., the magnetic term of $H_{tJ}$ in the basis (\ref{states})
has a four--fermion interaction $M_{ij} = (1 - n_i)(1 - n_j)$.

For the time being the state $\vert 0, \!\Downarrow\, \rangle_i$ will
be interpreted as a fictitious
``polarization state of the hole'', and not as the physical
doubly occupied state,
because we are interested in Eq.\ (\ref{tjeff}) only as an auxiliary
description of Eq.\ (\ref{tj}).  From this interpretation it follows that
the conserved quantity $G =\sum_{\sigma}N_{1\sigma}$
always corresponds to the number of electrons $N_{\rm el} = \sum_s{N_s}$
and $\delta=\sum_{\sigma}N_{0\sigma}/M=\sum_i n_i$ to
the doping. The extended model commutes with $\vec S$ and $\vec L$, and
from Eq.\ (\ref{conservation}) one easily sees that for any given
$G=N_{\rm el}$ the condition $N_{0\Downarrow}=0$ projects onto the sector
where pseudospin $L_z$ and total pseudospin $L$ attain
their maximum allowed value $L_z=L=(M-N_{\rm el})/2$.

In a grand--canonical approach the partition function of the
model (\ref{tjeff}) is
\begin{equation}
Z= {\rm Tr} \exp\{-\beta H_{\mu}\}, \quad
H_{\mu}=H_{tJ}^{\rm ex}-\sum_{\sigma} \mu^{0\sigma}N_{0\sigma},
\label{Z}
\end{equation}
where $\mu^{0\sigma}$ are the chemical potentials for the two species of
``holes''. In the basis (\ref{states})
a path--integral representation for $Z$ can be built--up immediately
by using Grassmann variables $\eta_i, \eta_i^*$
for the spinless fermions and {\it standard} SU(2)
coherent states $\vert {\vec \Omega}_i \rangle$ for the isospin
vectors. In fact, $\vec Q_i = \vec S_i + \vec L_i$ is the
sum of two {\it reducible} operators, for which vector addition does not
apply. The only eigenvalues of $Q_{i,\alpha}$ are
$q_z=\pm{1\over2}$ and indeed in the basis (\ref{states}) we can write
the isospin vector as a pure spin-${1\over2}$ operator:
$\vec Q_i = {1\over2}\vec\sigma_i$. Moreover, this property allows us to
introduce an expansion parameter, denoted $q$, by enlarging the dimensionality
of the SU(2) representation of $\vec Q_i$. For consistency we enlarge the
whole even sector (\ref{hub-spin},c) of su(2$\vert$2), setting
$S_{i,t}=q(1-n_i)$ and $L_{i,t}=qn_i$ in Eq.\ (\ref{hubbard_sf}).
Following the spin--wave approach adopted in Ref.\cite{last},
the generalization of $H_{\mu}$ at arbitrary $q$
is thus achieved by letting
\begin{eqnarray}
&& (P_{ij}-1)  \to  P_{ij}^{(-)} = 2(\vec Q_i \vec Q_j - q^2), \label{P} \\
&&  P_{ij}     \to  P_{ij}^{(+)} = 2(\vec Q_i \vec Q_j + q^2), \quad
\gamma_{i,\pm} \to q\pm Q_{i,z}.     \nonumber
\end{eqnarray}
Due to the presence of $\Delta_{ij}$ in Eq.\ (\ref{tjeff}),
the Grassmann integration over the variables $\eta_i^*$ and $\eta_i$
entering $Z$ can be performed {\it exactly} and we
obtain $Z=\int {\cal D}{\vec \Omega}\exp\{-S_{\rm eff}\}$,
where ${\cal D}{\vec \Omega}$ denotes the usual\cite{haldane} integration
over classical unit--vectors $\vec \Omega_i(\tau)$ and the effective lattice
action is
\begin{eqnarray}
\FL
 S_{\rm eff} =
&&
 {iq}\sum_i  \int_0^{\beta} d\tau {\vec {\cal A}_i}
                \partial_{\tau} {\vec \Omega}_i +
           {4q^2}{J\over2} \sum_{\langle ij \rangle} \int_0^{\beta} d\tau\,
                {\cal P}_{ij}^{(-)}
\label{eff_action} \\
&&
-  {\rm Tr}\ln \{[
                 \partial_{\tau} - 4q^2(\mu_i + \mu_{-} \Omega_{i,z})]
                   \delta_{ij} + 4q^2t\Lambda_{ij}{\cal P}_{ij}^{(+)}\}
\nonumber
\end{eqnarray}
with ${\vec {\cal A}_i}$ the Dirac potential\cite{haldane}
for a monopole of unit strength,
${\cal P}_{ij}^{(\pm)} = {1\over2}({\vec\Omega}_i {\vec\Omega}_j \pm 1)$,
and
\begin{equation}
\FL
 \mu_{\pm} = {1\over {2q}}{{\mu^{0\Uparrow}  \pm \mu^{0\Downarrow}}\over2},
\,\,\,\,
 \mu_i = \mu_{+} +
  {J\over2}\sum_k\Lambda_{ik}{\cal P}_{ik}^{(-)}.
\label{eff_chem_pot}
\end{equation}
The model (\ref{tjeff}) behaves as a correlated hopping, and in fact $\mu_i$
enters $S_{\rm eff}$ as a rotationally invariant local chemical potential
correlated to the background. In dealing with the extended model
$H_{tJ}^{\rm ex}$ itself, the
most natural choice is to treat the two species of holes equally,
thus setting $\mu_-=0$, whereas when considering the $t$-$J$ model one has to
impose the condition ${{\partial \ln Z}/{\partial \mu^{0\Downarrow}}}=0$.
As expected, in the latter case, the rotational symmetry in isospin space
is explicitly broken, as for $\mu_-$ is coupled to $\Omega_{i,z}$ and acts
as an effective magnetic field. However, as
discussed in Ref.\cite{last}, in the static
limit $t=0$ the ground--state energies of $H_{tJ}^{\rm ex}$ and
$H_{tJ}$ are degenerate and
we expect the degeneracy to remain for a physically sensible
range of the parameters $J/\vert t\vert$ and $\delta$. Hence, in this case,
one can investigate the zero temperature properties of the $t$-$J$ model
directly with the extended Hamiltonian $H_{tJ}^{\rm ex}$, thus working
with the simpler condition $\mu_-=0$.

The doping is accounted for by the third term in the r.h.s. of Eq.\
(\ref{eff_action}). This contribution is ineffective at zero doping, and
because for $\delta=0$ spin and isospin coincide (as for $\vec L_i=0$),
then $S_{\rm eff}$ correctly reduces to the standard lattice action of a
quantum antiferromagnet\cite{haldane}.
We also notice that, contrary to other proposed
functional integrals\cite{assa} for the $t$-$J$  model, the imaginary part
of Eq.\ (\ref{eff_action}) is a sum of standard Berry phases, so that
the quantization condition on the monopole flux\cite{haldane} is satisfied
for any $\delta$.
We remark that the classical field ${\vec \Omega}_i(\tau)$ is the
expectation value of the operator ${\vec Q}_i/q$, hence for $\delta\ne 0$
{\it physical} correlation functions can be evaluated only by differentiating
$Z$ with respect to suitable source terms, e.g.,
$H_{\rm s} = \sum_i {\vec h}_i {\vec S}_i \equiv
\sum_i {\vec h}_i(1-n_i){\vec Q}_i$, added to $H_{\mu}$.
Allowing for these modifications of Eq.\ (\ref{Z}), the effective action
is modified as well. However, the resulting Eq.\ (\ref{eff_action}) remains a
purely  classical expression, for which a gradient--expansion\cite{ale}
controlled by the parameter $q$ allows extraction of the relevant low--energy
behaviour.

The basis (\ref{states}) is suited to define an effective action
also for the Hubbard model. To this aim one only needs
to insert Eq.\ (\ref{op_electr}) into Eq.\ (\ref{hubbard}).
However, before performing this straightforward calculation, we
divide the lattice into two sublattices $A$ and $B$, and perform the
transformation
$c_{i,s}^{\dag} \to e^{i\theta_i} c_{i,s}^{\dag}$, where
$\theta_i=0$ for $i\in A$ and $\theta_i={\pi\over2}$ for $i\in B$,
so that ${H}_H$ is mapped into an equivalent Hamiltonian
${\hat H}_H$, whose hopping matrix satisfies to
$t{\hat \Lambda}_{ij}=-t{\hat \Lambda}_{ji}=t{\hat \Lambda}_{ji}^*$.
We make this change because it is known\cite{zhang} that the operator
$\vec M$, with components $M_{-}=\sum_i c^{\dag}_{i,\downarrow} \,
[(-)^{m_i}c^{\dag}_{i,\uparrow} +c_{i,\uparrow}]$, $M_+=M_-^{\dag}$, and
$M_z= \sum_i ({1\over2} - c^{\dag}_{i,\downarrow} c_{i,\downarrow})$,
where $m_i=0$ ($1$) for $i\in A$ ($B$), commutes with
$H = H_H + UL_z$.
By changing the phases we have $\vec M \to \vec Q$,
because the alternating sign $\epsilon_i=(-)^{m_i}$ cancels, and
${H}_H \to {\hat H}_H$, so that $[{\hat H}_H + UL_z, \vec Q]=0$. Hence,
this choice of the phases leads to the  representation of the
Hubbard model
\begin{equation}
{\hat H}_H = -t\sum_{i,j}{\hat\Lambda}_{ij} \left[ P_{ij} f_i^{\dag}f_j +
       (P_{ij}-1)C_{ij}\right] + H_{\rm at},
\label{hubeff}
\end{equation}
where $C_{ij}={1\over2}(f_if_j - f_j^{\dag}f_i^{\dag})$ and
$H_{\rm at} = U\sum_i \gamma_{i,-} n_i$, with a manifestly isospin invariant
hopping.
In Eq.\ (\ref{hubeff}) one easily identifies a term ${\hat H}_{0}$
conserving the number of objects of each species, and a
term ${\hat H}_{I}$ leading instead to transitions
$\Delta G = \pm 2$. Setting ${\hat \Lambda}_{ij} \to -\Lambda_{ij}$
via the transformation $f_i^{\dag} \to e^{i\theta_i} f_i^{\dag}$, we have
${\hat H}_{0} \to {\tilde H}_{0} =
H_{\rm at} + t\sum_{i,j}{\Lambda}_{ij} P_{ij} f_i^{\dag}f_j$, and
confronting this expression with the results in Ref.\cite{schad} one sees
that ${\tilde H}_{0}$ is also the particular bond--charge model
$H_{\rm bc} = H_H + t \sum_{i,j,s} \Lambda_{ij}c_{i,s}^{\dag}c_{j,s}
(n_{i,{\bar s}} + n_{j,{\bar s}})$. The equivalence of these different
looking models\cite{gros} is found because both have the same expression
${\tilde H}_{0} = H_{\rm at} - t\sum_{\langle i,j\rangle} P_{ij}^{01}$
in terms of abstract Hubbard operators, and holds because a global phase
change in one basis leads instead  to  site--dependent
transformations when rephrased in the other, e.g.,
\begin{equation}
f_i^{\dag} \to e^{i\vartheta} f_i^{\dag}  \,\,\,\,\leftrightarrow\,\,\,\,
c_{i,s}^{\dag} \to \exp\{i\vartheta(2n_{i,{\bar s}}-1)\}
c_{i,s}^{\dag}.
\label{phase}
\end{equation}
This is actually the main reason why, {\it before} making
use of Eq.\ (\ref{op_electr}), we have brought $H_H$ to a form with
antisymmetric adherence matrix. By replacing directly Eq. (\ref{op_electr})
into ${H}_H$ and $\vec M$, one obtains messy representations obscuring the
result  $[H_H + UL_z,\vec M]=0$.

Using Eq.\ (\ref{P}) and following the discussion leading to
Eq.\ (\ref{eff_action}), the Hamiltonian
${\hat H}_{\mu}={\hat H}_H - \mu N_{\rm el}$ can be
extended at arbitrary $q$ and its partition function represented as the
path--integral $Z=\int {\cal D}{\vec \Omega}\exp\{-S_{\rm eff}^H\}$,
\[
S_{\rm eff}^H = -4q^2\beta\mu_0 M +
                {iq}\sum_i  \int_0^{\beta} d\tau {\vec {\cal A}_i}
                \partial_{\tau} {\vec \Omega}_i  -
                {1\over2} \ln {\rm det}[K_{ij}^{\alpha\beta}],
\]
where  $\mu_0 = \mu/(2q)$ and, setting
$U_0 = U/(2q)$,  $h = \mu_0-{{U_0}/2}$,
\begin{eqnarray}
K_{ij}^{11} & = &
[{\partial}_{\tau} + 4q^2 ({U_0\over2} + h \,\Omega_{i,z})]\,\delta_{ij}
- 4q^2 t {\hat\Lambda}_{ij}{{\cal P}}_{ij}^{(+)},\nonumber \\
K_{ij}^{12} & = & -4q^2 t {\hat\Lambda}_{ij}{{\cal P}}_{ij}^{(-)},  \quad\,
K_{ij}^{21} =  -4q^2 t {\hat\Lambda}_{ij}{{\cal P}}_{ij}^{(-)}, \label{K} \\
K_{ij}^{22} & = &
[{\partial}_{\tau} - 4q^2 ({U_0\over2} + h \,\Omega_{i,z})]\,\delta_{ij}
- 4q^2 t {\hat\Lambda}_{ij}{{\cal P}}_{ij}^{(+)}.\nonumber
\end{eqnarray}
The square root of the determinant of the  block--matrix
$K_{ij}^{\alpha\beta}$ enters into the effective action
because\cite{drouffe}, due to the
presence of ${\cal C}_{ij} = {1\over2}(\eta_i\eta_j - \eta_j^*\eta_i^*)$ in
the classical counterpart of ${\hat H}_H$, we have performed the
integration with respect to the variables $\eta_i^{\alpha}$,
with $\eta_i^1=\eta_i$,
$\eta_i^2=\eta_i^*$, satisfying to the condition
$(\eta_i^{\alpha})^*= \eta_i^{\beta} \sigma_{x,{\beta\alpha}}$.
The Pfaffian structure of $S_{\rm eff}^H$ is quite natural: The off--diagonal
blocks vanish only for a ferromagnetic background, and setting
$\Omega_{i,\alpha}= \delta_{\alpha z}$ one can easily see that in this
case both effective actions $S_{\rm eff}$ and $S_{\rm eff}^H$ correctly
reduce to the logarithm of a free spinless fermion determinant.
At half--filling one has $h=0$, so that $S_{\rm eff}^H$ is rotationally
invariant in the isospin. The symmetry is instead explicitly broken away
from half--filling, as expected because of the relation
$q\Omega_{i,z}= \langle \vec\Omega_i \vert
S_{i,z} + q(1 - N_{{\rm el},i}) \vert \vec\Omega_i \rangle$,
where $N_{{\rm el},i} $ is the local electron number operator.

In general temporal fluctuations of the classical fields
${\vec \Omega}_i(\tau)$ (i.e., quantum fluctuations of the $\vec Q_i$'s)
are suppressed by letting $q\to \infty$. Assuming the generalization
(\ref{P}), in Eq.\ (\ref{op_electr}) one has
$c^{\dag}_{i,\uparrow} \propto q\pm Q_{i,z} $ and
$c^{\dag}_{i,\downarrow} \propto Q_{i,-}$, and at large isospin we
consistently\cite{klr} obtain $c^{\dag}_{i,\uparrow} = O(q)$ and
$c^{\dag}_{i,\downarrow} = O(\sqrt{q})$, so that the propagation of the
spin-$\downarrow$ electrons becomes smaller and eventually vanishes
at $q=\infty$. Hence, at mean--field level
$S_{\rm eff}^H$ reduces to the effective action $S_{\rm eff}^{\rm m.f.}$
that one would have obtained  by considering the Falikov--Kimball
Hamiltonian\cite{vollhardt,lieb}. This is a useful result, because one
can reasonably obtain sensible information about the Hubbard model
by systematically including corrections starting from a model, the
Falikov--Kimball, for which many exact results are known\cite{lieb}.

I am grateful to A. Muramatsu and S. Sorella for many stimulating discussions.
I also acknowledge useful discussions with A. Parola. This work is supported
by the Human Capital and Mobility program under contract $\#$ ERBCHBICT930475.


\begin{references}
\bibitem{three} Because $H^{(3)}$ is a permutational operator, a similar
                generalization can be applied also to the three--site term.
\end{references}
\end{document}